\documentclass[runningheads]{llncs}
\usepackage[T1]{fontenc}
\usepackage{amsmath}
\usepackage{makeidx}
\usepackage{cite}
\usepackage{multirow}
\usepackage{float}
\usepackage{graphicx}
\begin{document}
\title{Meta-Learning in Audio and Speech Processing: An End to End Comprehensive Review}
\titlerunning{Meta-Learning in Audio and Speech Processing Review}
%
\author{Athul Raimon\inst{1} \and Shubha Masti\inst{1} \and Shyam K Sateesh\inst{1} \and Siyani Vengatagiri\inst{1} \and  Bhaskarjyoti Das\inst{1}}

\authorrunning{Raimon et al.} 
%
\tocauthor{Athul Raimon, Shubha Masti, Shyam K Sateesh, Siyani Vengatagiri, Bhaskarjyoti Das}
\institute{%
  \textsuperscript{1}PES University, Bengaluru, India\\
  \texttt{\phantom{email}}
}

\maketitle              
\begin{abstract}
This survey overviews various meta-learning approaches used in audio and speech processing scenarios. Meta-learning is used where model performance needs to be maximized with minimum annotated samples, making it suitable for low-sample audio processing. Although the field has made some significant contributions, audio meta-learning still lacks the presence of comprehensive survey papers. We present a systematic review of meta-learning methodologies in audio processing. This includes audio-specific discussions on data augmentation, feature extraction, preprocessing techniques, meta-learners, task selection strategies and also presents important datasets in audio, together with crucial real-world use cases. Through this extensive review, we aim to provide valuable insights and identify future research directions in the intersection of meta-learning and audio processing.

\end{abstract}
\section{Introduction}\label{sec:intro}
Machine learning has made immense progress \cite{image_3} in recent years, mainly due to advancements in deep learning methods characterised by large labelled datasets. However, acquiring these large annotated datasets is often difficult due to the manual effort, cost, and privacy issues. In such cases, models trained using standard deep-learning methods often fail to perform well \cite{deepLearningNotGood}. Few-shot learning has emerged as a revolutionary technique in data-scarce scenarios \cite{notion_9}, aspiring to address these challenges by allowing models to learn from a few samples.

Few-shot learning (FSL) can be defined as the ability of a machine learning model to learn and generalise from a few training examples. Unlike most other learning scenarios, few-shot learning is usually performed on many tasks instead of one. This enables the model to acquire meta-knowledge that can be utilised when a few examples of a new task are provided. This approach is helpful in cases where it is impossible to gather an extensively annotated dataset, for instance, in the field of personalised medicine \cite{notion_17}, rare sound event classification \cite{notion_21}, or low-resource language processing \cite{notion_31,notion_33}.

Meta-learning \cite{image_8} is one of the most widely used techniques in few-shot learning. It aims to gain knowledge that can be easily applied to new problems with little data. Meta-learning algorithms do not emphasise learning a certain task; they learn to generalise across many different tasks.

This survey is organised as follows: Section 2 provides a background on meta-learning, detailing the foundational concepts necessary to understand the methodologies used in recent meta-learning approaches. Section 3 discusses audio-specific meta-learning approaches, including data preprocessing specifics, a glimpse of traditional FSL methods along with enhancements to these methods. It also delves into task selection, meta learners used and finally, use cases and common datasets.

\section{Background}\label{sec:abc}

Meta-learning or \textquotedblleft Learning to Learn\textquotedblright \cite{image_9} involves developing models that learn faster and with less data for new tasks using previous knowledge. Different from popular approaches of machine learning that require lots of data to train a model, meta-learning thrives in few-shot learning environments as it trains on different tasks to learn the best way to approach unseen classes. It is characterised by the query and support sets, losses to control the optimisation process, and encoders that convert raw data into appropriate features.

The query and support sets in meta-learning (collectively called Task Sets) define the few-shot learning scenario in training and testing. The support set is a small labelled set that provides the definition of the task for the model to learn. In an \(N\)-way \(K\)-shot context, the support set has \(N\) classes each containing \(K\) labelled examples. The model employs this basic information to acquire and develop specific tasks. The query set consists of unlabelled instances that belong to the same classes as the ones in the support set. It assesses the model’s performance once it has sharpened its knowledge using the support set. The model’s efficiency relies on how accurately it categorises the query set according to the information learnt from the support set.

Loss functions are essential components in the training of meta-learning models. They control the optimisation procedure by defining how closely the model's prognoses approximate the targets.

\begin{figure}
    \centering
    \includegraphics[width=1\linewidth]{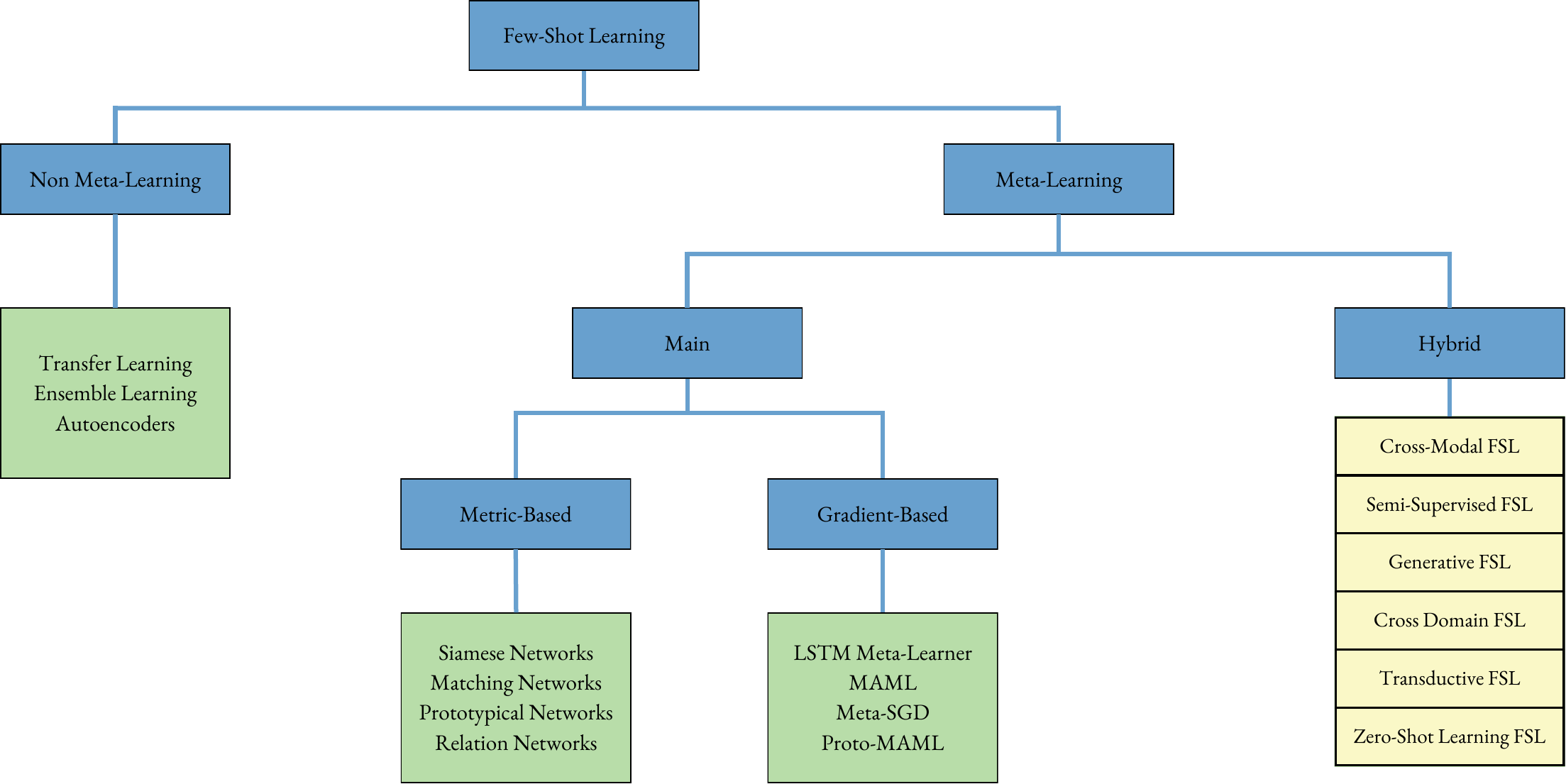}
    \caption{Overview of Few-Shot Learning Techniques}
\end{figure}

Metric-based meta-learning leverages distance metrics to classify data points by comparing them to labelled examples. The key idea is to learn an embedding space where similar instances are close together, and dissimilar ones are far apart. Prototypical networks (PN) \cite{image_21} or ProtoNets are one of the most popular metric-based techniques. They operate under the assumption that an embedding space exists, where samples from each class cluster around a single prototype representation. The goal is to learn and use this embedding space for few-shot classification. The accuracy of Prototypical Networks often depends on the choice of the embedding function used (for more details, refer to section 3.3).

Gradient-based meta-learning methods use gradient optimisation, allowing models to learn with little data and easily switch to other tasks using gradients previously learned in training. The goal is to find an initialisation or learning strategy that results in and enables fast adaptation by gradients. Model-Agnostic Meta-Learning (MAML) is a versatile meta-learning algorithm which operates by breaking down the dataset into tasks containing a support set and a query set. During training, MAML performs two main updates: an inner loop update, where model parameters are updated with gradient descent for each task and an outer loop update, where the updated parameters are used to compute the loss on the query set. The outer loop sums up the losses across tasks. The iterative process enables MAML to obtain a good initialization for fast adaptation. 

Dynamic Few-Shot Continual Learning (DFSL) \cite{DFSL_extra} can be considered a hybrid approach with metric and gradient-based meta-learning elements. It leans more towards the metric-based category due to its reliance on feature vectors and attention mechanisms to generate classification weights for novel classes. The classifier is first trained on abundant examples from the base classes to form robust feature representations and classification boundaries. It is then extended at inference time to recognise previously unseen (novel) classes based on a few labelled data. DFSL will thus incrementally expand the model to dynamically adapt to new classes without forgetting the old ones. 

\section{Audio Specific Meta-Learning Approaches}\label{sec:abc}

\subsection{Data Preprocessing}

\subsubsection{Sampling Rates.}
Sampling rate is the number of sound source samples played or recorded, usually measured in kilohertz (kHz) or cycles per second. A higher sample rate means more audio signal snapshots are captured, resulting in a more accurate digital representation of the original sound. In audio meta-learning, some of the more commonly observed sampling rates used in the surveyed papers are explained below.

\text{} \\
\noindent
\textit{44.1 kHz downsampled to 16 kHz.} This approach guarantees a high audio quality while being computationally efficient. Audio was generally mixed down to mono and resampled from 44kHz to 16kHz to minimise complexity and for faster processing of the data, as seen in \cite{notion_5,notion_10,notion_12,notion_13,notion_16,notion_33,notion_24}. 

\text{} \\
\noindent
\textit{44.1 kHz.} A standard sampling rate in audio applications that guaranteed high quality but was computationally expensive \cite{notion_7}.

\text{} \\
\noindent
We have observed that the downsampling from 44kHz to 16kHz is globally accepted in audio processing, especially in audio meta-learning.

\subsubsection{Features.}
Table~\ref{audio_features} explores the numerous features that can be extracted from an audio signal. These are important as they capture the temporal and spectral properties of the input signal.
\begin{table}[H]
    \centering
    \caption{Audio Features} \label{audio_features}
     \renewcommand{\arraystretch}{1.2} 
    \resizebox{0.9\textwidth}{!}{%
        \begin{tabular}{|p{4cm}|p{8.5cm}|}
            \hline
            \textbf{Feature} & \textbf{Description and Use Cases} \\
            \hline
            \hline
            Mel Spectrograms & They represent the power of an audio signal where the frequencies are on the Mel scale, closest to human ear perception. They were most used due to their ability to extract perceptually relevant features as shown in \cite{notion_3,notion_4,notion_12,notion_13,notion_24,notion_25,notion_33}.\\
            \hline
            Mel Filterbank & Extracted by taking the power spectrum of the audio signal and applying mel-scaled filters to it. They were employed for efficient representation as seen in \cite{notion_2,notion_26}. \\
            \hline
            Log Mel Filterbank & These features include applying 40 filters, each of which is Mel-scaled to the power spectrum, and then applying logarithm to the result to reduce the dynamic range. They were used in \cite{notion_6,notion_16}. \\
            \hline
            Log Mel Spectrogram & Mel spectrogram of the signal, augmented by taking the logarithm of the amplitude values to improve the dynamic range compression. They were useful for obtaining strong audio characteristics and was applied in \cite{notion_5,notion_7,notion_9,notion_22,notion_24}.\\
            \hline
            IS10 Acoustic Feature Set & Comprises 1582 dimensional acoustic features suitable for audio analysis. It extracted exhaustive audio features in \cite{notion_18}. \\
            \hline
            Spectrogram with Time-Frequency Masking & This involved using spectrograms coupled with T-F masking, which assisted in feature extraction by bringing out key spectral areas, seen in \cite{notion_29}. \\
            \hline
            SoundNet Conv5 & CNN aimed at audio tasks, and ConvNet 5 was one of its architectures used to obtain abstract representations of sounds. Used in \cite{notion_11}. \\
            \hline
            MIDI & Musical Instrument Digital Interface (MIDI) is applied to represent musical data digitally and was appropriate for tasks connected with music analysis, shown in \cite{notion_22}. \\
            \hline
        \end{tabular}%
    }
\end{table}

\noindent
Evidently, the usage of Mel Features is vast and reliable in audio meta-learning, consistently providing models with good features.

\subsubsection{Signal to Noise Ratio (SNR).}
SNR measures the level of a desired signal relative to the background noise level. A higher SNR during processing means greater clarity in the audio, while a lower SNR means more noise. SNR in meta-learning for audio is important as it can affect generalisability in noisy environments.

\text{} \\
\noindent
\textit{Dataset Augmentation with Noise.} Similar to \cite{notion_12}, \cite{notion_3} analyzed the performance of a model based on noise-augmented datasets for different SNR values as well. \cite{notion_12} added background noise to ESC-50, while \cite{notion_3} added white noise to FSD50k. These studies suggest that the models should be trained under various forms of noise to perform well in real-life conditions.
        
\text{} \\
\noindent
\textit{Robustness to Background Noise.} The approach proposed in \cite{notion_16} and \cite{notion_29} take forward the idea of how to make models more robust by explicitly considering background noise when training. In \cite{notion_16}, background noise was viewed as an additional class in creating an independent feature space that lessens the effect of false positives in a noisy environment. Similarly, \cite{notion_29} combined speech datasets with noise datasets and performed better than noise-aware traditional baselines.

\text{} \\
\noindent
\textit{SNR and Class Separation.}\cite{notion_16} and \cite{notion_3} also provided ways to control the SNR within the framework of class separation and polyphony. Samples of events with background noise were mixed to yield different SNRs in \cite{notion_16}, ensuring classes of events were separated well from background noise. An investigation was made into the impact of SNR and polyphony on few-shot learning, where matching the support set characteristics to testing data worked much better, especially in multi-label scenarios \cite{notion_3}. This paper shows that meta-learning for audio is very different from image classification and handles multi-label data and variation in SNR in a more complex manner.

\subsubsection{Data Augmentation Techniques.}
Data augmentation in audio processing can be considered critical in increasing the model's reliability and avoiding overfitting when working with a limited amount of data.

SpecAugment is an effective spectrogram data augmentation technique that performs masking on the input log-Mel spectrogram. A randomly selected time-frequency box within the input feature was zeroed out in \cite{notion_21}. Replacing groups of masks with the same value from masked positions encouraged variability, avoiding overfitting.

Inference-Time Data Augmentation is a strategy that decreased the work required for labelling samples since the selection window was moved by some amount of time, and new examples were created based on each example provided by the user. This increased the number of positive examples, improving the model's resilience without requiring further input from the human annotator \cite{notion_8}.

Mixup Augmentation implemented in \cite{notion_24}, formed new training examples by constraining the model to be linear between training examples, enhancing the model's generalisation on small datasets.

\subsection{Traditional FSL Methods}

\subsubsection{Prototypical Networks.}

\text{} \\
\noindent
\textit{Performance in Low-Data Scenarios.} Prototypical networks consistently outperform deep learning (DL) methods when the number of training samples ($n$) is less than 50. However, their performance tends to plateau and may not be as competitive as DL architectures when the training data is plentiful. Prototypical networks resist overfitting due to their strong distance-based classifier \cite{notion_9}.

\text{} \\
\noindent
\textit{Adaptations for Sound Event Detection.} \cite{notion_8} adapts few-shot learning approaches to automate acoustic event detection (AED), addressing open and closed-set problems. The adaptation to open set problems aims to enhance the efficiency of AED and reduce the need for manual labelling.

\text{} \\
\noindent
\textit{Distance Computation and Generalisation.} \cite{notion_6} highlights an approach in using prototypical networks for AED by computing the average distance of each query sample to every support sample, rather than averaging embeddings to create prototypes. This enhances generalization, reduces overfitting, and enables prototypical networks to outperform MAML and MetaOptNet.

\subsubsection{Dynamic Few-Shot Continual Learning (DFSL).}

\text{} \\
\noindent
\textit{Performance in Few-Shot Scenarios.}
DFSL adopts binary cross-entropy loss instead of categorical cross-entropy loss, along with global temporal pooling for weakly labelled data in the last convolutional layer for summarising the features \cite{notion_1}. Compared with the original classifier, the adapted DFSL can obtain a high F-score for novel classes and a slight performance loss for base classes. The DFSL variant based on cosine similarity indexing of base classes exhibited the best performance among all the baselines \cite{notion_3}. 

\text{} \\
\noindent
\textit{Incremental Learning and Knowledge Retention.} In incremental learning, DFSL is a popular method for classifying new sounds while retaining prior knowledge\cite{notion_15}. This is done by creating prototypes for each class using dynamic modules, bridging the gap between base and incremental sessions. Here, with the increase in the number of sessions, DFSL outperformed all baselines in terms of Average Accuracy and Performance Dropping rate.

\subsubsection{Model Agnostic Meta-Learning (MAML).}

\text{} \\
\noindent
\textit{Low-resource audio scenarios.}
To address the issue of limited medical data, MAML was trained using non-snoring sounds and tested on snoring sounds \cite{notion_17}, and the model performed well. It also adapted faster to new languages with small amounts of data \cite{notion_31}, showing improvements on other traditional multilingual pretraining approaches. When used to improve speech quality under various noisy environments \cite{notion_29}, it produced better outcomes with few samples.

\text{} \\
\noindent
\textit{Model-agnostic.} The model independence of MAML makes it flexible and capable of quickly transferring and learning from new tasks. It was used to predict the distribution of musical tokens \cite{notion_19} trained on a particular genre or composition. Results revealed that the genre-specific model was qualitatively superior to the other models, and the composition-specific model was superior in both qualitative and quantitative aspects.

\subsection{Enhancements to Traditional FSL Methods}
Meta-learning techniques for audio data differ from those for image data in several ways, primarily because of audio data's temporal and spectral properties. Audio-specific challenges demand that recent models, like MAML and PN, could be further developed to accommodate polyphony, low SNR environments, and multi-label classifications. Weakly labelled audio samples and rare audio events demand the development of robust models for smaller shots. This section will delve into these challenges and the enhancements that can be integrated to solve them.

\subsubsection{Changing Loss Function.}

\text{} \\
\noindent
\textit{Attentional Similarity for Segment-Level Focus.} \cite{notion_12} suggests using attentional similarity to emphasise a certain segment in the audio clip, where it calculates segment-by-segment similarity using feature maps that capture temporal dependencies of inputs and using rank-1 approximation.

\text{} \\
\noindent
\textit{Hierarchical and Multi-Label Classification.} Hierarchical relationships between labels can be exploited to improve model generalization. LaD-ProtoNet \cite{notion_7} designed the loss function to prioritise parent labels over child labels. Hierarchical Prototypical Networks \cite{notion_20} were also used to optimise the model using hierarchical cross-entropy loss where each level of the tree was considered an independent multi-class classification task.

\text{} \\
\noindent
\textit{Episodic Triplet Mining.} \cite{notion_13} uses episodic triplet mining (ETM) as a loss function, comparing distances within triplets to stabilise the model, demonstrating significant performance improvements over traditional loss models.

\text{} \\
\noindent
\textit{Multi-Step Loss for Stability and Convergence.} A multi-step loss (MSL) \cite{notion_33} procedure was used to resolve the issues of vanishing and exploding gradients. The losses calculated at each step in the inner loop of MAML were combined with a weighted importance vector to stabilise the training process and speed up convergence.

\text{} \\
\noindent   
\subsubsection{Encoders.}
Encoders take raw data and turn them into usable features for meta-learning models. The choice of the encoder is the driving factor behind how well models generalise examples with few shots, especially in metric-based methods like prototypical networks.

\text{} \\
\noindent
\textit{Standard CNNs.} CNNs are popular due to their efficiency in extracting raw features from the audio data. They also capture hierarchical patterns of input features, with multiple layers of convolutions, activations, and pooling \cite{notion_6,notion_8,notion_12,notion_16}. 

\text{} \\
\noindent
\textit{VGG-Based Encoders.} VGGs are built upon the convolutional network to achieve higher embedding accuracy by modifying its architecture. VGG-11 \cite{notion_2} takes the average of outputs from frames, and VGG-M \cite{notion_10} adds more layers for finer detail extraction. 

\text{} \\
\noindent
\textit{Long Short Term Memory Networks (LSTM).} Single-layer LSTMs were used to average per-frame outputs into clip embeddings \cite{notion_2}. However, VGG-based encoders outperformed them and proved better in learning meta-audio patterns. 

\text{} \\
\noindent
\textit{ResNet Architectures.} ResNet-34 \cite{notion_10}, adapted for input spectrograms, using convolutional layers and residual connections. It outperforms VGG-M due to its higher parameter count that captures intricate audio patterns effectively \cite{notion_15}.

\subsubsection{Hybrid FSL Methods.}
Hybrid FSL models modify pre-existing meta-learning models, such as MAML and prototypical networks, among other optimisation strategies, to make the learning process more adaptive and increase performance on data-scarce tasks like few-shot learning. This also helps achieve better generalisation and faster convergence.  Table~\ref{maml}, Table~\ref{prototypical} and Table~\ref{hybrid} outline the same.

\begin{table}[H]
    \centering
    \caption{MAML and its Derivatives} \label{maml}
    \label{table:audio_features}
     \renewcommand{\arraystretch}{1.2} 
    \resizebox{1\textwidth}{!}{%
        \begin{tabular}{|p{4cm}|p{9cm}|}
            \hline
            \textbf{Method} & \textbf{Description} \\
            \hline
            \hline
            Almost No Inner Loop (ANIL)& Simplification of MAML where inner loop updates are made only for the final layer of the network \cite{notion_22}. Gives the same level of performance with less complexity.\\
            \hline
            Speaker Adaptation with Modified MAML& Specifically adapts modules according to their function in the model \cite{notion_32}, achieving better results across different metrics.\\
            \hline
            Mixture Density Network-Based Meta-Learning&Combines mixture density networks with meta-learning, so that it is agnostic to the model used \cite{notion_26}; focuses on quick adaptation with little data.\\
            \hline
            Task-Adaptive Parameter Transformation (TAPT)& The model is dynamically initialised \cite{notion_25} using gradients of the initial randomised parameters. It combines task-specific and task-wide knowledge, resulting in faster convergence and better performance.\\\hline
        \end{tabular}%
    }
\end{table}

\begin{table}[H]
    \centering
    \caption{Prototypical Networks and its Derivatives} \label{prototypical}
    \label{table:audio_features}
     \renewcommand{\arraystretch}{1.2} 
    \resizebox{1\textwidth}{!}{%
        \begin{tabular}{|p{4cm}|p{9cm}|}
            \hline
            \textbf{Method} & \textbf{Description} \\
            \hline
            \hline
            MetaOptNet & A feature-space meta-learning classifier built upon a linear SVM instead of a nearest neighbor-based approach. The model outperforms the baseline when using the ResNet-12 architecture but is computationally heavy \cite{notion_6,notion_22}.\\
            \hline
            Prototypical Networks& Enhanced by techniques like maximising inter-class distance and minimising intra-class distance to avoid overfitting \cite{notion_23}, improving performance.\\
            \hline
        \end{tabular}%
    }
\end{table}

\begin{table}[H]
    \centering
    \caption{Other Hybrid Approaches} \label{hybrid}
    \label{table:audio_features}
     \renewcommand{\arraystretch}{1.2} 
    \resizebox{1\textwidth}{!}{%
        \begin{tabular}{|p{4cm}|p{9cm}|}
            \hline
            \textbf{Method} & \textbf{Description} \\
            \hline
            \hline
            Task-Adaptive Module and Transductive Propagation Network (TPN)& Learn from test data during training. They improve feature representation and classification accuracy through task-specific extraction and label propagation \cite{notion_21}.\\
            \hline
            Hierarchical Prototypical Network (HPN) & Enhances audio classification by leveraging hierarchical information \cite{notion_20}, treating each level of a hierarchical tree as an independent multi-class classification task, leading to superior performance compared to traditional Prototypical Networks.\\
            \hline
 Attentional GNNs&Use an attention mechanism to weigh examples in the support set differently \cite{notion_11}, improving performance by focusing on the most relevant samples.\\\hline
        \end{tabular}%
    }
\end{table}

\subsection{Task Selection}
\subsubsection{Open and Closed Sets.}
In general, closed-set problems are those where both training and testing include a fixed number of pre-defined classes ($C$), and they are commonly used in the traditional few-shot learning setting. Open-set problems cope with a variable number of classes for training and testing, i.e., previously unseen classes are introduced. This is a typical scenario for sound event detection: detecting a target sound from a sequence of unknown, previously unheard sounds from an unbounded number of classes. 

\cite{notion_8} explores few-shot learning for open-set problems. They propose using labelled examples of a target sound and finding the occurrence of the target sound within a long recording. The novel strategies introduce automatically constructed labelled negative examples and data augmentation at inference time to improve detection accuracy and minimise the labelling effort. The average AUPRC of the method was commendable in detecting unseen target keywords, with an average of only five labelled examples used.

\subsubsection{One vs Rest.}
Every classifier is trained to differentiate between one class and the rest, thus minimising the number of interactions between multiple labels and simplifying the learning process. The sampling strategy in this technique is such that each classifier is trained one class at a time to make the learning process easier by focusing on a subset of the support set \cite{notion_5}. The created subsets are based on the number of labels attached to the query example. To tackle multi-label classification \cite{notion_7}, each sample is converted to multiple single-label tasks. Prototypes are constructed by calculating the average of each class's embeddings. Episodes include a query set and a support set consisting of multiple classes. The labels for each subset contain the positive class, and N-1 randomly sampled negative classes \cite{notion_5}. In total, \(N \times K\) examples are chosen for the support set \cite{notion_7}. The query example is classified by calculating the distance between the query and the prototypes in the embedding space, producing a probability distribution over the classes.

\subsubsection{Domain of Query.}
The term Domain of Query relates to the context in which a model is built using different datasets, languages, or types of sounds. It is important in meta-learning since it determines how well models transfer learned information from training data to test data. Domain Mismatch is always associated with reduced performance; thus, effective techniques must be employed to address such fluctuations.

\text{} \\
\noindent
\textit{Impact of Domain Mismatch.}  All the meta-learning methods in \cite{notion_6} reflected a decrease in performance when trained and tested on different domains and was even more prominent in animal sounds than music. Prototypical Networks tend to overfit to the training domain, which makes generalising across different domains difficult.

\text{} \\
\noindent
\textit{Use of External Datasets for Training.} For snore sound recognition \cite{notion_17}, external image and audio datasets were used for meta-training, while the snoring dataset (MPSSC \cite{MPSSC}) was used for meta-testing. MAML identified features from the image data, resulting in a UAR of 41\% even though the datasets were dissimilar. When using mel-spectrogram features for training, UAR increased to 60.2\%, showing MAML's efficiency on external datasets.  

\text{} \\
\noindent
\textit{Automatic Speech Recognition (ASR).} A generalisation problem was found in ASR \cite{notion_31} when they trained on three languages and tested on a different language. Meta-ASR, which is MAML, was employed to update the parameters for the model instead of each language head of the model. This enhanced generalisation on other languages and reduced the Character Error Rate (CER) scores for all different languages used in testing . 

\text{} \\
\noindent
\textit{Comparative Analysis of Meta-Learning Methods.} Five primary datasets and two meta-testing datasets were used in \cite{notion_4}. These datasets had both fixed and variable-length audio clips. Meta-curvature (Gradient-Based) outperformed other baselines. It was also proved that in-domain meta-learning methods outperform transfer learning from external datasets as joint training across domains resulted in degradation of performance.

\subsubsection{Uniform sampling.}
In meta-learning, effective sampling strategies are important to train models that are generalisable using tasks with few examples of data involved. For this reason, sampling affects the distribution and diversity of data samples at training time and, consequently, directly influences how well a model can learn to distinguish between classes.

\text{} \\
\noindent
\textit{Balanced Pairwise Sampling.}
This technique keeps a constant ratio of noise and other classes in the datasets, balancing the pairs and further separating the noise and event classes. This leads to better detection of target events. In \cite{notion_16}, this technique was implemented between target classes and noise samples for few-shot detection.

\text{} \\
\noindent
\textit{Consistent Class Ratios.}
Uniform sampling keeps all classes in an equal proportion with respect to the parent class, making the learning process more balanced and effective. \cite{notion_20} consisted of other sampling methods, such as parent sampling, where all classes in the child set were selected from a single parent and random sampling, one achieved by simply using a random split of classes. In the end, uniform sampling outperformed both parent and random sampling in accuracy and F1 score, showing the effectiveness of consistency in class ratio to enhance model performance.

\subsection{Meta Learners Used.}

\subsubsection{Convolutional Neural Networks (CNNs).} Convolution network serves as a relation module \cite{notion_15}, optimised using the SGD algorithm. Snoring sounds were also recognised using CNNs \cite{notion_17} with a meta-learning approach where inner and outer learning rates were specified as 0.01 and 0.001, respectively. In the benchmark  \cite{notion_4}, a lightweight hybrid CRNN was used, suggesting a bias toward CNN-based structures in audio classification problems.

\subsubsection{Attentional Meta Learners.} Attentional graph neural networks (GNN) \cite{notion_11} have been used for few-shot audio classification. Attention-masked transformer architectures outperformed LSTMs  \cite{notion_19}, which solidified the claim that attention can be used for sequential data, such as MIDI files.

\subsubsection{Other Architectures.} Gaussian Mixture Models (GMMs) have been utilised to implement speaker identification \cite{notion_26}, where the mixture density network (MDN) mapped input features to the GMM's parameters and gave an overall probability density function for the features. A TTS model was made based on FastSpeech 2 \cite{notion_32} for text-to-speech applications. Results in \cite{notion_29} demonstrated that the U-Net architecture could also be useful in restoring degraded audio signals with limited data for few-shot speech enhancement. 

\text{} \\
\noindent
Adam was the most used optimiser in the reviewed papers due to its adaptive learning rate properties. Adam was employed for snore sound recognition with a CNN \cite{notion_17}, in a C-RNN for few-shot audio classification \cite{notion_4}, in a U-Net for few-shot speech enhancement \cite{notion_29}, for low-resource speech recognition with a CNN \cite{notion_33}, and in the outer loop for musical meta-learning with transformers \cite{notion_19}.

\subsection{Use Cases and Datasets}
Table~\ref{datasets} describes the most prevalent use cases and datasets seen in audio meta-learning. 

\begin{table}[H]
    \centering
    \caption{Common Datasets and Use Cases} \label{datasets}
    \label{table:dataset}
    \renewcommand{\arraystretch}{1.2} 
    \resizebox{0.9\textwidth}{!}{%
        \begin{tabular}{|p{3cm}|p{6cm}|p{3cm}|}
            \hline
            \textbf{Dataset} & \textbf{Description} & \textbf{Use Case(s)} \\
            \hline
            \hline
            ESC-50 \cite{esc50} & Environment Sound & Classification \cite{notion_22} \\
            \hline
            AudioSet \cite{audioSet} & Audio Event classes from YouTube (Hierarchical graph) & Classification \cite{notion_11} \\
            \hline
            VoxCeleb \cite{voxCeleb} & Short clips of human speech from YouTube interviews & Classification \cite{notion_2} \\
            \hline
            FSD50k \cite{fsd50k} & Sound Events (sampled from AudioSet) & Classification \cite{notion_3} \\
            \hline
            VCTK \cite{VCTK}& English speech corpus & Classification \cite{notion_13}, Generation \cite{notion_32} \\
            \hline
            Lakh MIDI \cite{lakhMIDI} & Songs in MIDI format & Generation \cite{notion_19} \\
            \hline
            Common Voice \cite{commonVoice} & Multi-language dataset speech corpus & Generation \cite{notion_33} \\
            \hline
            TIMIT \cite{timit} & Acoustic-Phonetic Continuous Speech Corpus & Enhancement \cite{notion_29} \\
            \hline
        \end{tabular}%
    }
\end{table}

\section{Conclusion}
This survey describes major use cases and appropriate datasets for audio-specific meta-learning, focusing on the importance of sampling rates and popular input features. Data augmentation techniques add to model robustness in a complementary way. Hybrid models usually exhibit better performance over the baseline meta-learning models, especially if enriched with an effective loss function, attention similarities, and advanced sampling strategies. Meta-learners like CNNs, attentional GNNs, and U-Nets optimized using Adam reflect their effectiveness for a wide range of audio tasks. The spectral-temporal nature of the audio data calls for further modifications to the basic design of the meta-learning models, which were originally envisioned for the image domain. We trust that this survey paper provides an in-depth introduction and overview to researchers in the field, aiding in the advancements in the creation of superior models in data-scarce audio scenarios.

\bibliographystyle{splncs04}
\bibliography{ref}

\end{document}